\pdfoutput=1
\documentclass[11pt, acmsmall,nonacm]{acmart}
\usepackage{fullpage}
\usepackage{graphicx}
\usepackage{etoolbox}
\apptocmd{\thebibliography}{\raggedright}{}{}
\usepackage{pdfsync}
\usepackage{multicol}
\usepackage{geometry}
\usepackage{titlesec}
\usepackage{datasheet}
\usepackage{xspace}
\usepackage[T1]{fontenc}
\usepackage{url}
\usepackage{hyperref}
\usepackage{cleveref}
\usepackage{enumitem}
\usepackage[normalem]{ulem}
\usepackage{soul}

\newcommand{\edit}[1]{#1}

\setlist{nolistsep}

\begin{document}

\title{Datasheets for Datasets}

\author{Timnit Gebru}
\affiliation{%
  \institution{\edit{Black in AI}}
}
\author{Jamie Morgenstern}
\affiliation{%
  \institution{\edit{University of Washington}}
}
\author{Briana Vecchione}
\affiliation{%
  \institution{Cornell University}
}
\author{Jennifer Wortman Vaughan}
\affiliation{%
  \institution{Microsoft Research}
}
\author{Hanna Wallach}
\affiliation{%
  \institution{Microsoft Research}
}
\author{Hal Daum\'{e} III}
\affiliation{%
  \institution{Microsoft Research;}
  \institution{University of Maryland}
}
\author{Kate Crawford}
\affiliation{%
  \institution{\edit{Microsoft Research}}
}

\renewcommand{\shortauthors}{Gebru et al.}

\maketitle

\section{Introduction}
Data plays a critical role in machine learning. Every machine learning
model is trained and evaluated using data, quite often in the
form \edit{of} static dataset\edit{s}. The characteristics of these
datasets fundamentally influence a model's behavior: \edit{a} model is
unlikely to perform well in the wild if its deployment context does
not match its training or evaluation datasets, or if these datasets
reflect unwanted
\edit{societal} biases. Mismatches like this can have especially severe
consequences when machine learning \edit{models are} used in
high-stakes domains\edit{,} such as criminal
justice~\cite{facerec_law_us, facerec_law_elsewhere, compass},
hiring~\cite{hiring_ml_us}, critical infrastructure~\cite{water,
grid}, \edit{and} finance~\cite{newinvestor}. \edit{E}ven in other
domains, mismatches may lead to loss of revenue or public relations
setbacks. Of particular concern are recent examples showing that
machine learning models can reproduce or amplify unwanted societal
biases reflected in training \edit{datasets} \cite{gendershades,
amazon, word2vec}. For these and other reasons, the World Economic
Forum suggests that all entities should document the provenance,
creation, and use of machine learning datasets in order to avoid
discriminatory outcomes~\cite{wef}.

Although data provenance has been studied extensively in the databases
community~\cite{provenance,datahub}, it is rarely discussed in the
machine learning community\edit{. D}ocumenting the creation and use of
datasets has received even less attention. Despite the importance of
data to machine learning, there is \edit{currently no} standardized
process for documenting machine learning datasets.

To address this gap, we propose \emph{datasheets for datasets}. In the
electronics industry, every component, no matter how simple or
complex, is accompanied with a datasheet describing its operating
characteristics, test results, recommended usage, and other
information. By analogy, we propose that every dataset be accompanied
with a datasheet that documents its motivation, composition,
collection process, recommended uses, and so on. Datasheets for
datasets have the potential to increase transparency and
accountability within the machine learning community, mitigate
unwanted \edit{societal} biases in machine learning \edit{models},
facilitate greater reproducibility of machine learning results, and
help researchers and practitioners \edit{to} select more appropriate datasets
for their chosen tasks.

After outlining our objectives below, we describe the process by which
we developed \edit{datasheets for datasets}. \edit{We then provide a set of
questions designed to elicit the information that a datasheet for a
dataset might contain, as well as a} workflow for dataset creators to
use when answering these questions. We conclude with a summary of the
impact \edit{to date} of datasheets for datasets and a discussion of
implementation challenges and avenues for future work.

\label{sec:intro}

\subsection{Objectives}
\label{sec:objectives}
Datasheets for datasets are intended to address the needs of two key
stakeholder groups: dataset creators and dataset consumers. For
dataset creators, the primary objective is to encourage careful
reflection on the process of creating, distributing, and maintaining a
dataset, including any underlying assumptions, potential risks or
harms, and implications of use. For dataset consumers, the primary
objective is to ensure they have the information they need to make
informed decisions about using a \edit{dataset.} Transparency on the
part \edit{of} dataset creators is necessary for dataset consumers to
be sufficiently well informed that they can select appropriate
datasets for their
\edit{chosen} tasks and avoid unintentional misuse.\footnote{\edit{We note that in
some cases, the people creating a datasheet for a dataset may not be
the dataset creators, as was the case with the example datasheets that
we created as part of our development process.}}

Beyond these two key stakeholder groups, datasheets \edit{for datasets} may
\edit{be} valuable to policy makers, consumer advocates, \edit{investigative
journalists}, individuals whose data is included in datasets, and
\edit{individuals} who may be impacted by models trained or evaluated \edit{using} datasets. They also serve a secondary objective of facilitating
greater reproducibility of machine learning results: researchers and
practitioners \edit{without access to a dataset may be able to use the
information in its datasheet to create alternative datasets with
similar characteristics.}

Although we provide a set of questions designed to elicit the
information that a datasheet for a dataset might contain, \edit{these
questions} are not intended to be prescriptive. Indeed, we expect that
datasheets will \edit{necessarily} vary depending on factors such as the
domain or existing organizational infrastructure and workflows. For
example,
\edit{some the questions are appropriate for academic researchers publicly
releasing datasets for the purpose of enabling future research, but
less relevant for product teams creating internal datasets for
training proprietary models.  As another example,}~\citet{statements}
outline a proposal similar to datasheets for
datasets \edit{specifically intended for language-based
datasets. Their} questions may \edit{be} naturally integrated into a
datasheet for a language-based dataset as appropriate.

We emphasize that the process of creating a datasheet is not intended
to be automated. Although automated documentation processes are
convenient, they run counter to our objective of encouraging dataset
creators to carefully reflect on the process of creating,
distributing, and maintaining a dataset.

\section{Development Process}
\label{sec:process}
We refined the \edit{questions and workflow provided in the next
section} over a period of roughly \edit{two years},
incorporating \edit{many rounds of feedback}.

First, leveraging our own experiences as researchers with diverse
backgrounds working in different domains and institutions, we drew on
our knowledge of dataset characteristics, unintentional misuse,
unwanted \edit{societal} biases, and other issues to produce an initial set
of questions \edit{designed to elicit information about these topics. We}
then ``tested'' the\edit{se} questions by creating example datasheets for two
\edit{widely used} datasets: Labeled Faces in the Wild~\cite{lfw} and Pang
and Lee's polarity dataset~\cite{polarity}. \edit{We chose these
datasets in large part because their creators provided exemplary
documentation, allowing us to easily find the answers to many of the
questions.} While creating the\edit{se example} datasheets,
we \edit{found} gaps in \edit{the} questions, as well as redundancies
and lack of clarity. We \edit{therefore} refined \edit{the} questions
and distributed them to product teams in two major US-based technology
companies, in some cases helping \edit{teams} to create datasheets for
their datasets and observing where \edit{the} questions did not
achieve their intended objectives. Contemporaneously, we circulated an
initial draft of this paper to colleagues through social media and on
arXiv (draft posted 23 March 2018). Via these channels we received
extensive comments from dozens of researchers, practitioners,
and \edit{policy makers}. We also worked with \edit{a team of lawyers}
to review the questions from a legal perspective.

We incorporated this feedback to yield the questions and workflow
\edit{provided} in the next section\edit{:} We \edit{added and removed}
questions,
\edit{refined the content of the} questions\edit{,} and \edit{reordered the questions}
to better match the key stages of the dataset lifecycle. Based on our
experiences with product teams, we reworded the questions to
discourage yes/no answers, added a section on ``Uses\edit{,''} and
deleted a section on ``Legal and Ethical Considerations.'' We found
that product teams were more likely to answer questions about legal
and ethical considerations if they were integrated into sections about
the relevant stages of the dataset
\edit{lifecycle} rather than grouped together. \edit{Finally, f}ollowing feedback
from \edit{the team of lawyers}, we removed questions \edit{that} explicitly
ask\edit{ed} about compliance with regulations, and introduced factual
questions intended to elicit relevant information about compliance
without requiring dataset creators to make legal judgments.

\section{Questions and Workflow}
\label{sec:workflow_uses}
In this section, we provide a set of questions \edit{designed to elicit} the
information that a datasheet for a dataset might contain, as well as a
workflow for dataset creators to use when answering these
questions. The questions are grouped into sections that roughly match
the key stages of the dataset lifecycle: motivation, composition,
collection process, preprocessing/cleaning/labeling, uses,
distribution, and maintenance. This grouping encourages dataset
creators to reflect on the process of creating, distributing, and
maintaining a dataset, and even alter this process in response \edit{to}
their reflection.  We note that not all questions will be applicable
to all datasets; those that do not apply \edit{should} be skipped.

To illustrate how these questions might be answered in practice, we
provide in the appendix \edit{an} example datasheet for Pang and Lee's
polarity dataset~\cite{polarity}. \edit{We answered some of the
questions with ``Unknown to the authors of the datasheet.'' This is
because we did not create the dataset ourselves and could not find the
answers to these questions in the available documentation. For an
example of a datasheet that was created by the creators of the
corresponding dataset, please see that of Cao and
Daum\'e}~\cite{cao2019toward}.\footnote{See \url{https://github.com/TristaCao/into_inclusivecoref/blob/master/GICoref/datasheet-gicoref.md}.} \edit{We
note that even dataset creators may be unable to answer all of the
questions provided in this section. We recommend answering as many
questions as possible rather than skipping the datasheet creation
process entirely.}

\subsection{Motivation}

The questions in this section are primarily intended to encourage
dataset creators to clearly articulate their reasons for creating the
dataset and to promote transparency about funding interests.
The latter may be particularly relevant for datasets created for
research purposes.\\

\begin{itemize}

\item \textbf{For what purpose was the dataset created?} Was there a specific task in mind? Was there a specific gap that needed to be filled? Please provide a description.

\item \textbf{Who created the dataset (e.g., which team, research group) and on behalf of which entity (e.g., company, institution, organization)?}

\item \textbf{Who funded the creation of the dataset?} If there is an associated grant, please provide the name of the grantor and the grant name and number.

\item \textbf{Any other comments?}

\end{itemize}

\subsection{Composition}

Dataset creators should read through \edit{these questions} prior to
any data collection and then provide answers once \edit{data} collection is
complete. Most of the questions \edit{in this section} are intended to
provide dataset consumers with the information they need to make
informed decisions about using the dataset for their chosen
tasks. Some of the questions are \edit{designed to elicit} information
about compliance with the EU's General Data Protection Regulation
(GDPR) or comparable regulations in other jurisdictions.

\edit{Questions that apply only to datasets that relate to people are
grouped together at the end of the section. We recommend taking a
broad interpretation of whether a dataset relates to people. For
example, any dataset containing text that was written by people
relates to people.}\\

\begin{itemize}

\item \textbf{What do the instances that comprise the dataset
    represent (e.g., documents, photos, people, countries)?} Are there
  multiple types of instances (e.g., movies, users, and ratings;
  people and interactions between them; nodes and edges)? Please
  provide a description.

\item \textbf{How many instances are there in total (of each type, if appropriate)?}

\item \textbf{Does the dataset contain all possible instances or is it
    a sample (not necessarily random) of instances from a larger set?}
  If the dataset is a sample, then what is the larger set? Is the
  sample representative of the larger set (e.g., geographic coverage)?
  If so, please describe how this representativeness was
  validated/verified. If it is not representative of the larger set,
  please describe why not (e.g., to cover a more diverse range of
  instances, because instances were withheld or unavailable).

\item \textbf{What data does each instance consist of?} ``Raw'' data
  (e.g., unprocessed text or images) or features? In either case,
  please provide a description.

\item \textbf{Is there a label or target associated with each
    instance?} If so, please provide a description.

\item \textbf{Is any information missing from individual instances?}
  If so, please provide a description, explaining why this information
  is missing (e.g., because it was unavailable). This does not include
  intentionally removed information, but might include, e.g., redacted
  text.

\item \textbf{Are relationships between individual instances made
    explicit (e.g., users' movie ratings, social network links)?} If
  so, please describe how these relationships are made explicit.

\item \textbf{Are there recommended data splits (e.g., training,
    development/validation, testing)?} If so, please provide a
  description of these splits, explaining the rationale behind them.

\item \textbf{Are there any errors, sources of noise, or redundancies
    in the dataset?} If so, please provide a description.

\item \textbf{Is the dataset self-contained, or does it link to or
    otherwise rely on external resources (e.g., websites, tweets,
    other datasets)?} If it links to or relies on external resources,
    a) are there guarantees that they will exist, and remain constant,
    over time; b) are there official archival versions of the complete
    dataset (i.e., including the external resources as they existed at
    the time the dataset was created); c) are there any restrictions
    (e.g., licenses, fees) associated with any of the external
    resources that might apply to a \edit{dataset consumer}? Please provide
    descriptions of all external resources and any restrictions
    associated with them, as well as links or other access points, as
    appropriate.

\item \textbf{Does the dataset contain data that might be considered
    confidential (e.g., data that is protected by legal privilege or
    by doctor\edit{--}patient confidentiality, data that includes the content
    of individuals' non-public communications)?} If so, please provide
    a description.

\item \textbf{Does the dataset contain data that, if viewed directly,
    might be offensive, insulting, threatening, or might otherwise
    cause anxiety?} If so, please describe why.

\end{itemize}

\edit{If the dataset does not }relate to people, you may skip the remaining questions in this section.

\begin{itemize}

\item \textbf{Does the dataset identify any subpopulations (e.g., by
    age, gender)?} If so, please describe how these subpopulations are
  identified and provide a description of their respective
  distributions within the dataset.

\item \textbf{Is it possible to identify individuals (i.e., one or
    more natural persons), either directly or indirectly (i.e., in
    combination with other data) from the dataset?} If so, please
    describe how.

\item \textbf{Does the dataset contain data that might be considered
    sensitive in any way (e.g., data that reveals rac\edit{e} or ethnic
    origins, sexual orientations, religious beliefs, political
    opinions or union memberships, or locations; financial or health
    data; biometric or genetic data; forms of government
    identification, such as social security numbers; criminal
    history)?} If so, please provide a description.

\item \textbf{Any other comments?}

\end{itemize}

\subsection{Collection Process}

As with the \edit{questions in the} previous section, dataset creators should
read through these questions prior to any data collection to flag
potential issues and then provide answers once collection is complete.
\edit{In addition to the goals outlined in the previous section, the
questions in this section are designed to elicit information that may
help researchers and practitioners to create alternative datasets with
similar characteristics. Again, questions that apply only to datasets
that relate to people are grouped together at the end of the
section.}\\

\begin{itemize}

\item \textbf{How was the data associated with each instance
    acquired?} Was the data directly observable (e.g., raw text, movie
  ratings), reported by subjects (e.g., survey responses), or
  indirectly inferred/derived from other data (e.g., part-of-speech
  tags, model-based guesses for age or language)? If \edit{the} data was reported
  by subjects or indirectly inferred/derived from other data, was the
  data validated/verified? If so, please describe how.

\item \textbf{What mechanisms or procedures were used to collect the
    data (e.g., hardware apparatus\edit{es} or sensor\edit{s}, manual human
    curation, software program\edit{s}, software API\edit{s})?} How were these
    mechanisms or procedures validated?

\item \textbf{If the dataset is a sample from a larger set, what was
    the sampling strategy (e.g., deterministic, probabilistic with
    specific sampling probabilities)?}

\item \textbf{Who was involved in the data collection process (e.g.,
    students, crowdworkers, contractors) and how were they compensated
    (e.g., how much were crowdworkers paid)?}

\item \textbf{Over what timeframe was the data collected?} Does this
  timeframe match the creation timeframe of the data associated with
  the instances (e.g., recent crawl of old news articles)?  If not,
  please describe the timeframe in which the data associated with the
  instances was created.

\item \textbf{Were any ethical review processes conducted (e.g., by an
    institutional review board)?} If so, please provide a description
  of these review processes, including the outcomes, as well as a link
  or other access point to any supporting documentation.

\end{itemize}

\edit{If the dataset does not relate to people, you may skip the remaining questions in this section.}

\begin{itemize}

\item \textbf{Did you collect the data from the individuals in
    question directly, or obtain it via third parties or other sources
    (e.g., websites)?}

\item \textbf{Were the individuals in question notified about the data
    collection?} If so, please describe (or show with screenshots or
  other information) how notice was provided, and provide a link or
  other access point to, or otherwise reproduce, the exact language of
  the notification itself.

\item \textbf{Did the individuals in question consent to the
    collection and use of their data?} If so, please describe (or show
  with screenshots or other information) how consent was requested and
  provided, and provide a link or other access point to, or otherwise
  reproduce, the exact language to which the individuals consented.

\item \textbf{If consent was obtained, were the consenting individuals
    provided with a mechanism to revoke their consent in the future or
    for certain uses?} If so, please provide a description, as well as
  a link or other access point to the mechanism (if appropriate).

\item \textbf{Has an analysis of the potential impact of the dataset
    and its use on data subjects (e.g., a data protection impact
    analysis) been conducted?} If so, please provide a description of
  this analysis, including the outcomes, as well as a link or other
  access point to any supporting documentation.

\item \textbf{Any other comments?}

\end{itemize}

\subsection{Preprocessing/cleaning/labeling}

Dataset creators should read through these questions prior to any
preprocessing, cleaning, or labeling and then provide answers once
these tasks are complete. The questions in this section are intended
to provide dataset consumers with the information they need to
determine whether the ``raw'' data has been processed in ways that are
compatible with their chosen tasks. For example, text that has been
converted into a ``bag-of-words'' is not suitable for tasks involving
word order.\\

\begin{itemize}

\item \textbf{Was any preprocessing/cleaning/labeling of the data done
    (e.g., discretization or bucketing, tokenization, part-of-speech
    tagging, SIFT feature extraction, removal of instances, processing
    of missing values)?} If so, please provide a description. If not,
  you may skip the remain\edit{ing} questions in this section.

\item \textbf{Was the ``raw'' data saved in addition to the preprocessed/cleaned/labeled data (e.g., to support unanticipated future uses)?} If so, please provide a link or other access point to the ``raw'' data.

\item \textbf{Is the software \edit{that was} used to preprocess/clean/label the \edit{data} available?} If so, please provide a link or other access point.

\item \textbf{Any other comments?}

\end{itemize}

\subsection{Uses}

\edit{The} questions \edit{in this section} are intended to encourage dataset
creators to reflect on the tasks for which the dataset should and
should not be used. By explicitly highlighting these tasks, dataset
creators can help dataset consumers to make informed decisions,
thereby avoiding potential risks or harms.\\

\begin{itemize}

\item \textbf{Has the dataset been used for any tasks already?} If so, please provide a description.

\item \textbf{Is there a repository that links to any or all papers or systems that use the dataset?} If so, please provide a link or other access point.

\item \textbf{What (other) tasks could the dataset be used for?}

\item \textbf{Is there anything about the composition of the dataset or the way it was collected and preprocessed/cleaned/labeled that might impact future uses?} For example, is there anything that a \edit{dataset consumer} might need to know to avoid uses that could result in unfair treatment of individuals or groups (e.g., stereotyping, quality of service issues) or other \edit{risks or} harms (e.g., \edit{legal risks,} financial harms\edit{)?} If so, please provide a description. Is there anything a \edit{dataset consumer} could do to mitigate these \edit{risks or} harms?

\item \textbf{Are there tasks for which the dataset should not be used?} If so, please provide a description.

\item \textbf{Any other comments?}

\end{itemize}

\subsection{Distribution}

Dataset creators should provide answers to these questions prior to
distributing the dataset either internally within the entity on behalf
of which the dataset was created or externally to third parties.\\

\begin{itemize}

\item \textbf{Will the dataset be distributed to third parties outside of the entity (e.g., company, institution, organization) on behalf of which the dataset was created?} If so, please provide a description.

\item \textbf{How will the dataset will be distributed (e.g., tarball on website, API, GitHub)?} Does the dataset have a digital object identifier (DOI)?

\item \textbf{When will the dataset be distributed?}

\item \textbf{Will the dataset be distributed under a copyright or other intellectual property (IP) license, and/or under applicable terms of use (ToU)?} If so, please describe this license and/or ToU, and provide a link or other access point to, or otherwise reproduce, any relevant licensing terms or ToU, as well as any fees associated with these restrictions.

\item \textbf{Have any third parties imposed IP-based or other restrictions on the data associated with the instances?} If so, please describe these restrictions, and provide a link or other access point to, or otherwise reproduce, any relevant licensing terms, as well as any fees associated with these restrictions.

\item \textbf{Do any export controls or other regulatory restrictions apply to the dataset or to individual instances?} If so, please describe these restrictions, and provide a link or other access point to, or otherwise reproduce, any supporting documentation.

\item \textbf{Any other comments?}

\end{itemize}

\subsection{Maintenance}

As with the \edit{questions in the} previous section, dataset creators
should provide answers to these questions prior to distributing the
dataset. The questions \edit{in this section} are intended to
encourage dataset creators to plan for dataset maintenance and
communicate this plan \edit{to} dataset consumers.\\

\begin{itemize}

\item \textbf{Who \edit{will be} supporting/hosting/maintaining the dataset?}

\item \textbf{How can the owner/curator/manager of the dataset be contacted (e.g., email address)?}

\item \textbf{Is there an erratum?} If so, please provide a link or other access point.

\item \textbf{Will the dataset be updated (e.g., to correct labeling
    errors, add new instances, delete instances)?} If so, please
  describe how often, by whom, and how updates will be communicated to
  \edit{dataset consumers} (e.g., mailing list, GitHub)?

\item \textbf{If the dataset relates to people, are there applicable
    limits on the retention of the data associated with the instances
    (e.g., were \edit{the} individuals in question told that their data would
    \edit{be} retained for a fixed period of time and then deleted)?} If so,
    please describe these limits and explain how they will be
    enforced.

\item \textbf{Will older versions of the dataset continue to be
    supported/hosted/maintained?} If so, please describe how. If not,
  please describe how its obsolescence will be communicated to \edit{dataset
  consumers}.

\item \textbf{If others want to extend/augment/build on/contribute to
    the dataset, is there a mechanism for them to do so?} If so,
  please provide a description. Will these contributions be
  validated/verified? If so, please describe how. If not, why not? Is
  there a process for communicating/distributing these contributions
  to \edit{dataset consumers}? If so, please provide a description.

\item \textbf{Any other comments?}

\end{itemize}

\section{Impact and Challenges}
\label{sec:limitations}
Since circulating an initial draft of this paper in March 2018,
datasheets for datasets have already gained traction in a number of
settings. Academic researchers have adopted our proposal and released
datasets with accompanying datasheets~\cite[e.g.,][]{recipeqa2018,
dataset1, choi2018quac, daume20gicoref}. Microsoft, Google, and IBM
have begun to pilot datasheets for datasets internally within product
teams. Researchers at Google published follow-up work on \emph{model
cards} that document machine learning models~\cite{model_cards} and
released a \emph{data card} (a lightweight version of a datasheet)
\edit{along} with the Open Images dataset~\cite{openimages}. Researchers at
IBM proposed \emph{factsheets}~\cite{factsheets} that document various
characteristics of AI services, including whether the datasets used to
develop the services are accompanied with datasheets. \edit{The Data
Nutrition Project incorporated some of the questions provided in the
previous section into the latest release of their Dataset Nutrition
Label}~\cite{nutritionsecondgen}. Finally, the Partnership on AI, a
multi-stakeholder organization focused on \edit{studying and
formulating} best practices for developing and deploying \edit{AI
technologies}, is working on industry-wide documentation guidance that
builds \edit{on} datasheets\edit{ for datasets}, model cards, and
factsheets.\footnote{\url{https://www.partnershiponai.org/about-ml/}}

These initial successes have also revealed implementation challenges
that may need to be addressed to support wider adoption. Chief among
them is the need for dataset creators to modify the questions and
workflow \edit{provided in the previous} section based on their existing
organizational infrastructure and workflows. We \edit{also note
that} \edit{the} questions and workflow
may \edit{pose} \edit{problems} for dynamic datasets. If a dataset
changes only infrequently, we recommend accompanying updated versions
with updated datasheets.

Datasheets for datasets do not provide a complete solution to
mitigating unwanted \edit{societal} biases or potential risks or
harms. Dataset creators cannot anticipate every possible use of a
dataset, and identifying unwanted \edit{societal} biases often requires
additional labels indicating demographic information \edit{about}
individuals, which may not be available to dataset creators for
reasons including those individuals' data protection and
privacy~\cite{holstein2018improving}.

When creating datasets that relate to people, \edit{and hence their
accompanying datasheets,} it may be necessary for dataset creators to
work with experts in other domains such as anthropology\edit{, sociology,
and science and technology studies.} There are complex and contextual
social, historical, and geographical factors that influence how best
to collect data \edit{from} individuals \edit{in a manner that is respectful}.

Finally, creating datasheets for datasets will necessarily impose
overhead on dataset creators. Although datasheets may reduce the
amount of time that dataset creators spend answering one-off questions
about datasets, the process of creating a datasheet will always take
time, and organizational infrastructure and workflows\edit{---not to mention
incentives---}will need to be modified to accommodate this investment.

\edit{Despite these implementation} challenges, there are many benefits to
creating datasheets for datasets. In addition to facilitating better
communication between dataset creators and dataset consumers,
datasheets provide \edit{an} opportunity for dataset creators to distinguish
themselves as prioritizing transparency and
accountability. Ultimately, we believe that the benefits to the
machine learning community outweigh the costs.

\subsection*{Acknowledgments}
\label{sec:acknowledgments}
We thank Peter Bailey, Emily Bender, Yoshua Bengio, Sarah Bird, Sarah Brown,
Steven Bowles, Joy Buolamwini, Amanda Casari, Eric Charran, Alain
Couillault, Lukas Dauterman, Leigh Dodds, Miroslav Dud\'{i}k, Michael
Ekstrand, No\'emie Elhadad, Michael Golebiewski, Nick Gonsalves,
Martin Hansen, Andy Hickl, Michael Hoffman, Scott Hoogerwerf, Eric
Horvitz, Mingjing Huang, Surya Kallumadi, Ece Kamar, Krishnaram
Kenthapadi, Emre Kiciman, Jacquelyn Krones, Erik Learned-Miller,
Lillian Lee, Jochen Leidner, Rob Mauceri, Brian Mcfee, Emily
McReynolds, Bogdan Micu, Margaret Mitchell, Sangeeta Mudnal, Brendan O'Connor, Thomas
Padilla, Bo Pang, Anjali Parikh, Lisa Peets, Alessandro Perina,
Michael Philips, Barton Place, Sudha Rao, Jen Ren, David Van Riper, Anna Roth,
Cynthia Rudin, Ben Shneiderman, Biplav Srivastava, Ankur Teredesai,
Rachel Thomas, Martin Tomko, Panagiotis Tziachris, Meredith Whittaker,
Hans Wolters, Ashly Yeo, Lu Zhang, and the attendees of the
Partnership on AI's April 2019 ABOUT ML workshop for valuable
feedback.

\bibliographystyle{ACM-Reference-Format}
\bibliography{sample-base}

\pagebreak
\appendix
\section{Appendix}
\begin{figure*}[p]
  \includegraphics[page=1,width=1.0\textwidth]{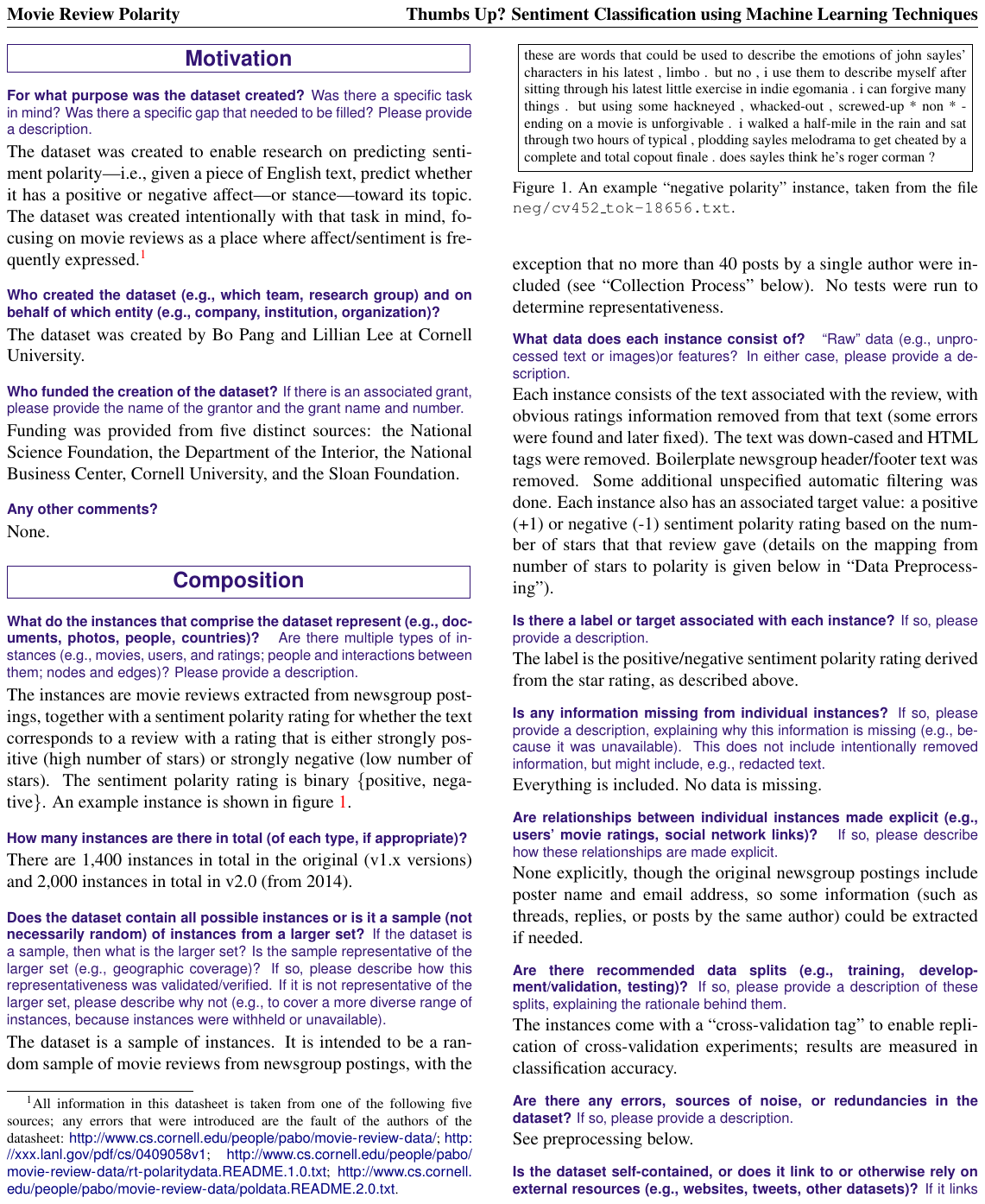}
  \caption{\label{fig:polarity1}Example datasheet for Pang and Lee's
    polarity dataset~\cite{polarity}, page 1.}
  \vspace{-1cm}
\end{figure*}

\begin{figure*}[p]
  \includegraphics[page=4,width=1.0\textwidth]{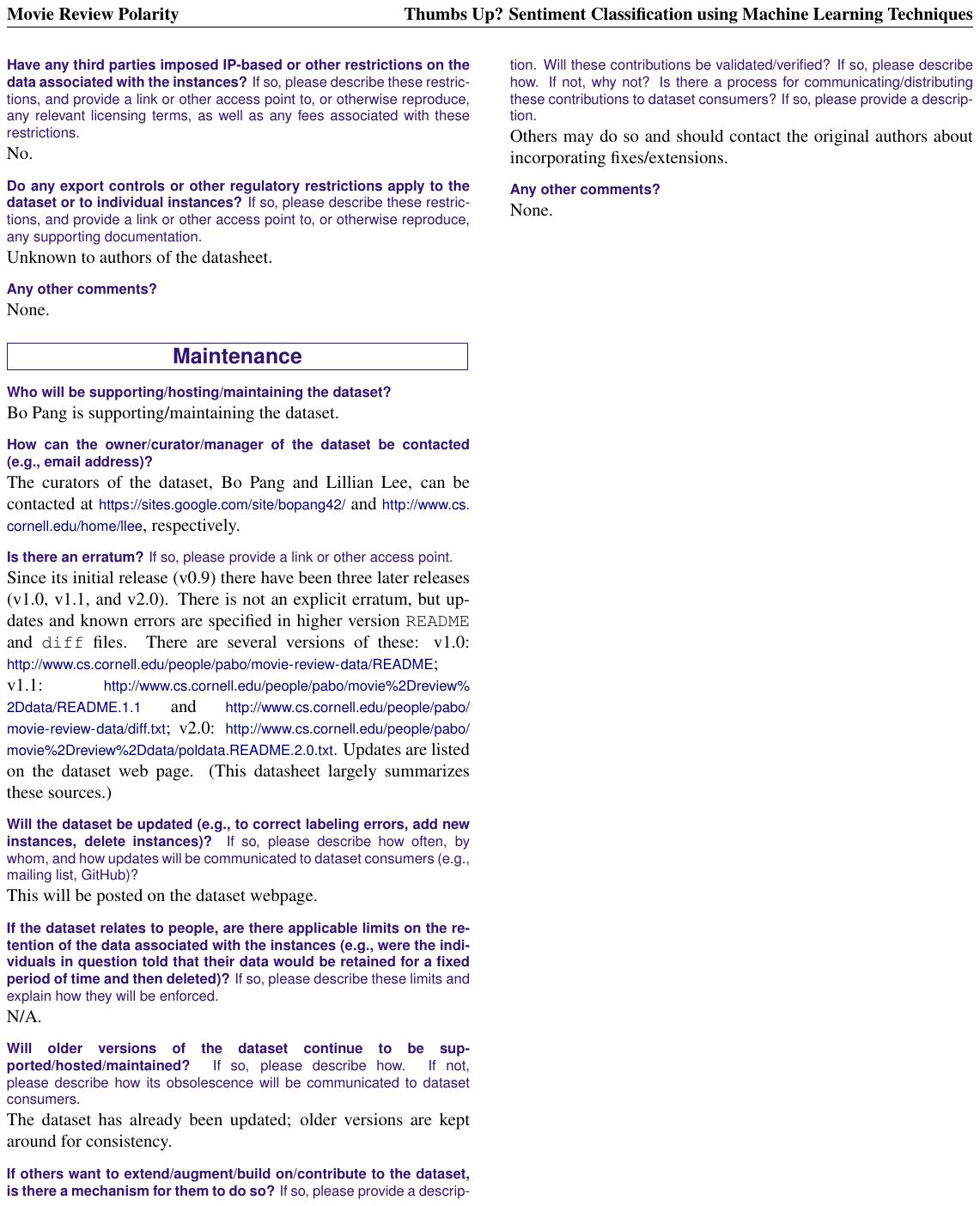}
  \vspace{-1cm}
  \caption{\label{fig:polarity4}Example datasheet for Pang and Lee's
  polarity dataset~\cite{polarity}, page 4.}
\end{figure*}

In this appendix, we provide \edit{an example datasheet for}
Pang and Lee's
polarity dataset~\cite{polarity} (figure~\ref{fig:polarity1} to
figure~\ref{fig:polarity4}).

\begin{figure*}[p]
  \includegraphics[page=2,width=1.0\textwidth]{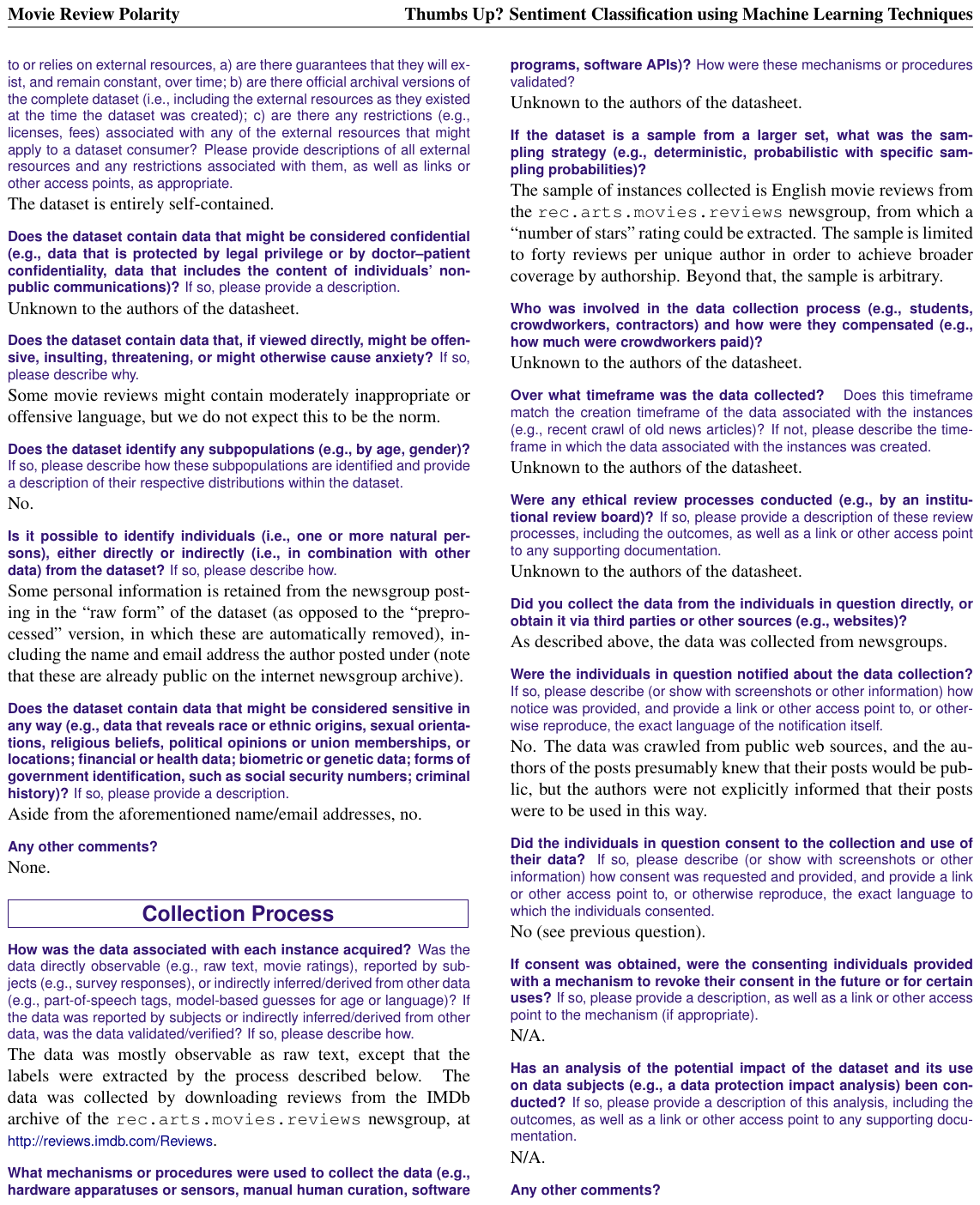}
  \caption{\label{fig:polarity2}Example datasheet for Pang and Lee's
    polarity dataset~\cite{polarity}, page 2.}
  \vspace{-1cm}
\end{figure*}

\begin{figure*}[p]
  \includegraphics[page=3,width=1.0\textwidth]{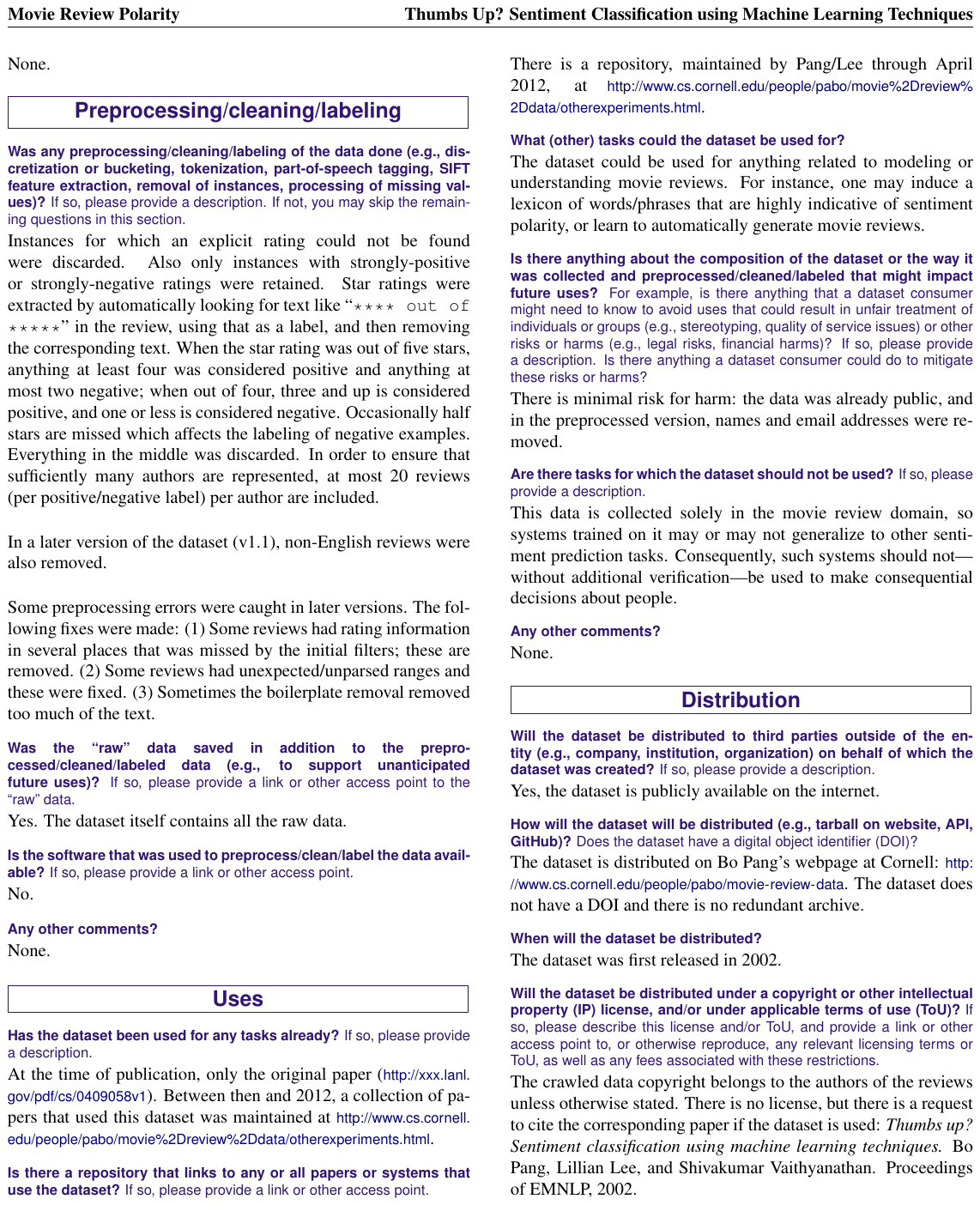}
  \vspace{-1cm}
  \caption{\label{fig:polarity3}Example datasheet for Pang and Lee's
  polarity dataset~\cite{polarity}, page 3.}
\end{figure*}

\end{document}